\documentclass[preprint,5p,authoryear]{elsarticle}
\input{preamble}
\newcommand{\Proj}{\mathcal{P}}

\newcommand{\NN}{\mathcal{N}}

\newcommand{\VV}{\mathcal{V}}

\newcommand{\RR}{\mathbb{R}}

\newtheorem{Proposition}{Proposition}
\newtheorem{Assumption}{Assumption}
\newtheorem{Remark}{Remark}
\newtheorem{Lemma}{Lemma}
\newtheorem{Theorem}{Theorem}
\newtheorem{Definition}{Definition}
\allowdisplaybreaks
\begin{document} \begin{sloppypar}\begin{frontmatter}
\title{Collision-Free Bearing-Driven Formation Tracking for Euler–Lagrange Systems
%Bearing-based Formation Tracking Control of Multiple Euler-Lagrange Systems
}
%\author{Haoshu Cheng, Martin Guay, Shimin Wang, Yunhong Che
%\thanks{%This work was supported in part by NSERC.
%}
\author[NTU]{Haoshu~Cheng}
\author[QUKOC]{Martin~Guay}
\author[MIT]{Shimin~Wang}
\author[MIT]{Yunhong Che}

\address[NTU]{Nanyang Technological University, Singapore 639798}
\address[QUKOC]{Queen's University, Kingston, ON K7L 3N6, Canada}
\address[MIT]{Massachusetts Institute of Technology, Cambridge,  MA 02142, USA}

\tnotetext[footnoteinfo]{This research was supported by NSERC.  \\
H.~Cheng is with the School of Electrical and Electronic Engineering, Nanyang Technological University, Singapore 639798.\\
%Martin Guay is with Queen's University, Kingston, ON K7L 3N6, Canada (e-mail: guaym@queensu.ca).\\
Martin Guay is with 
Queen's University, Kingston, ON K7L 3N6, Canada (martin.guay@queensu.ca).\\
Shimin Wang and Yunhong Che are with Massachusetts Institute of Technology, Cambridge, MA 02142, USA(bellewsm@mit.edu, yunhche@mit.edu).\\
Corresponding author: Shimin Wang.}% \\
%(Corresponding author: Shimin Wang).}

%}
%\thanks{}
%\maketitle

\begin{abstract}
In this paper, we investigate the problem of tracking formations driven by bearings for heterogeneous Euler–Lagrange systems with parametric uncertainty in the presence of multiple moving leaders. To estimate the leaders’ velocities and accelerations, we first design a distributed observer for the leader system, utilizing a bearing-based localization condition in place of the conventional connectivity assumption. This observer, coupled with an adaptive mechanism, enables the synthesis of a novel distributed control law that guides the formation towards the target formation, without requiring prior knowledge of the system parameters. Furthermore, we establish a sufficient condition, dependent on the initial formation configuration, that ensures collision avoidance throughout the formation evolution. The effectiveness of the proposed approach is demonstrated through a numerical example.
\end{abstract}

\end{frontmatter}
\section{Introduction}\label{sec:intro}

Formation control is a fundamental problem in the cooperative control of multi-agent systems, aiming to achieve a desired geometric pattern through a distributed control law \citep*{alfriend2009spacecraft,defoort2008sliding,wang2024distributed}.%formation2020%wang2018adaptive%
Due to its broad range of applications, including search and rescue, environmental monitoring, and surveillance, this topic has garnered significant attention in the control and systems community \citep*{beard2006UAVs,wang2025finite}. 
 %
%Among various formation control methods, bearing-based formation control is a prominent approach in which the inter-agent bearings serve as the controlled variables \cite{trinh2018bearing}.%formation2020
Among the various formation control problems, the tracking of formations driven by bearing constraints has not been sufficiently investigated. The approach to addressing this problem is commonly referred to as bearing-based formation control~\citep*{zhao2015bearing}.

The development of bearing-based formation control is grounded in the theories of bearing rigidity and bearing-based localization, which were rigorously established in \cite{zhao2015bearing} and \cite{zhao2016localizability}, respectively. 
Subsequently, an extended version of bearing rigidity incorporating the unit quaternion formalism was proposed in \cite{michieletto2019formation}, accompanied by a corresponding control design.  
The  Bearing-Ratio-of-Distance (B-RoD) rigidity framework was introduced in \cite{cao2019bearing} and applied to formation control.
Other advances include the consideration of double-integrator dynamics with damping in \cite{tron2018bearing}, and the formation with desired bearings satisfying the persistent excitation property in \cite{tang2021formation}. 
%a velocity-estimation-based scheme for formation tracking control in \cite{su2022bearing}
Besides, the effect of external disturbances was further examined in \cite{trinh2021robust}. Additional related studies can be found in \cite{tang2022relaxed} and the references therein.

%Despite these advances, most existing results focus on simple agent dynamics, such as single- or double-integrator models, which are insufficient for capturing the behavior of robotic systems commonly employed in industrial applications. 
%
Despite this rich body of work, most existing studies focus on idealized kinematic or simplified dynamic models, such as single- or double-integrator systems \citep*{mesbahi2010graph}.%formation2020
While such models allow elegant theoretical analysis, they fail to capture the nonlinear, coupled dynamics, actuator constraints, and parameter uncertainties inherent in practical robotic systems \citep*{spong2020robot,wang2018adaptive}.
%
%This discrepancy between theoretical models and realistic systems has been widely recognized \cite{zavlanos2007distributed,wang2025finite}, motivating the study of bearing-based formation control for more realistic dynamics.
In reality, many robotic platforms, such as UAVs, manipulators, and underwater vehicles, are governed by Euler–Lagrange dynamics with uncertain parameters arising from factors like friction, payload variations, and unmodeled dynamics \citep*{spong2020robot}.
From a practical standpoint, it is therefore both important and necessary to investigate bearing-based formation control for Euler–Lagrange systems, which provide a more realistic modelling framework for robotic systems \citep*{wang2025finite,wang2022leaderless}.

Furthermore, collision avoidance is a fundamental safety requirement in dense formations, yet it has not been adequately addressed alongside formation maintenance in the bearing-based setting.
%To date, only a few studies have addressed this challenge. 
%
Practical missions such as cooperative transportation, environmental monitoring, and search-and-rescue often involve multiple moving leaders coordinating the formation. However, most existing works consider only static target formations \citep*{zhao2023adaptive,li2020adaptive}.
Specifically, \cite{zhao2023adaptive} investigates bearing-based formation control for Euler-Lagrange systems under a leader-follower structure, while \cite{li2020adaptive} addresses the leaderless case. 
{In addition, both approaches}  do not ensure inter-agent collision avoidance.
%

% \cite{li2020adaptive} and %extend the line of research in \cite{li2020adaptive,zhao2023adaptive} by proposing
In this paper, we propose a new control design for bearing-based formation control of Euler-Lagrange systems, featuring three key innovations. 
First, we design a distributed observer under the bearing-based localization condition with multiple designated leaders, rather than relying on a standard connectivity condition, and assuming a single leader. 
This localization condition ensures both the uniqueness of the target formation and convergence to the state of the leader system. 
Second, by integrating the distributed observer with the adaptive control technique, we develop a novel distributed control law that allows Euler-Lagrange agents with unknown parameters to achieve the target formation. 
Finally, we derive a sufficient condition, based on the initial configuration, that guarantees collision avoidance between agents.
The proposed framework addresses key challenges, including tracking leaders with time-varying velocities generated by linear systems, multi-leader coordination under bearing-based localization rather than conventional connectivity assumptions, and collision avoidance amid the nonlinear dynamics of Euler–Lagrange systems. 
%
%To the best of our knowledge, this integrated approach has not been addressed in the bearing-based formation control literature.
%
% Compared to the existing results in~\cite{li2020adaptive,zhao2023adaptive}, our formulation presents a more challenging setting in at least three key aspects. 
%  %
%  First, it accommodates leaders with a time-varying velocity generated by a linear system, whereas prior works are restricted to static leaders. 
%   %
%  Second, to enable tracking of a moving target formation, the distributed observer must operate under multiple leaders and rely on bearing-based localization conditions, rather than the more conventional connectivity assumptions commonly used in cooperative problems of multi-agent systems. 
%   %
%  Third, due to the strong nonlinearity inherent in Euler–Lagrange systems, ensuring inter-agent collision avoidance poses a significant challenge. 
%   %
%  To the best of our knowledge, this problem has not yet been addressed within the context of bearing-based formation control for Euler–Lagrange systems.

The remainder of this paper is organized as follows: Section~\ref{sec:Pre} presents preliminaries and formulates the problem. Section~\ref{sec three} shows the main theoretical results. Section~\ref{sec four} provides a numerical example to illustrate the proposed design and verify its effectiveness.

\noindent {\em Notation.}   The notation $\|x\|$ represents the Euclidean norm of the vector $x$, while $\|A\|$ indicates the spectral norm of the real matrix $A$. For column vectors $x_i, \, i = 1, \dots, s$, $$\mathrm{col}(x_1, \dots, x_s) = [x_1^T, \dots, x_s^T]^T.$$ $\mathbf{1}_n$ denotes an $n$-dimensional column vector with all entries 1, while $\mathbf{0}$ denotes a matrix of appropriate dimensions filled with zeros. $I_n$ represents the identity matrix of size $n$. $\otimes$ denotes the Kronecker product. For a real symmetric matrix $A \in \RR^{w \times w}$, let $\lambda_{\min}(A)$ and $\lambda_{\max}(A)$ denote the minimum and maximum eigenvalues, respectively, while the eigenvalues of \( A \) are ordered as $\lambda_1(A) \geq \lambda_2(A) \geq \cdots \geq \lambda_w(A)$.
      For a vector $x= \mbox{col}(x_{1},\ldots,x_{s}) \in \RR^{s}$, $$\textnormal{diag}(x)=\begin{bmatrix}
	x_{1} & & \\ & \ddots & \\&& x_{s}
\end{bmatrix}.$$ For matrices $A_{1},\ldots,A_{s}$, $$\textnormal{blkdiag}(A_{1},\ldots,A_{s})=\begin{bmatrix}
	A_{1}&& \\ &\ddots& \\ && A_{s}
\end{bmatrix}.$$
   $\textnormal{rand}(x, y) \in \mathbb{R}^{x \times y}$ denotes a stochastic matrix with uniformly distributed elements in the interval $(0, 1)$.

\section{Preliminaries and Problem Formulation}\label{sec:Pre}

\subsection{Graph Theory and Formation}
In this paper, we consider the problem of tracking formations driven by bearings for \( n~(n \geq 3) \) agents in \( d~(d \geq 2) \) dimensions. The index set of the agents is denoted by \( \mathcal{V} = \{1, \ldots, n\} \), with \( \mathcal{V}_{l} = \{1, \ldots, n_{l}\} \) and \( \mathcal{V}_{f} = \{n_{l}+1, \ldots, n\} \) representing the sets of \( n_{l} \) leaders and \( n_{f} = n - n_{l} \) followers, respectively. The interaction among agents is modelled by a static undirected graph \( \mathcal{G} = (\mathcal{V}, \mathcal{E}) \), where \( \mathcal{E} \subseteq \mathcal{V} \times \mathcal{V} \) denotes the edge set consisting of \( 2m \) directed edges. If \( (i, j) \in \mathcal{E} \), then agent \( j \) is said to be a neighbor of agent \( i \), and the neighbor set of agent \( i \) is defined as \( \mathcal{N}_{i} = \{ j \mid (i, j) \in \mathcal{E} \} \). Since the graph is undirected, \( (i, j) \in \mathcal{E} \) implies \( (j, i) \in \mathcal{E} \).
 
  The head point of agent \( i \) at time \( t \) is denoted by \( q_{i}(t) \in \mathbb{R}^{d} \). Let the stacked position vector of all agents be defined as \[ q(t) = \mathrm{col}(q_{l}(t), q_{f}(t)) \in \mathbb{R}^{dn}, \] 
  where 
\begin{align*}
	q_{l}(t) &= \mathrm{col}(q_{1}(t), \ldots, q_{n_{l}}(t)) \in \mathbb{R}^{dn_{l}}\\
	q_{f}(t) &= \mathrm{col}(q_{n_{l}+1}(t), \ldots, q_{n}(t)) \in \mathbb{R}^{dn_{f}}.
\end{align*}
  For simplicity, we define the relative position between agents \( i \) and \( j \) as \( q_{ij}(t) = q_{i}(t) - q_{j}(t) \).
  
Given the relative position vector \( q_{ij}(t) \), we define the bearing vector from agent \( i \) to agent \( j \) as
\begin{align*}
    g_{ij}(t) = \dfrac{q_{ij}(t)}{\|q_{ij}(t)\|}.
\end{align*}
For a given bearing vector \( g_{ij}(t) \), the associated orthogonal projection matrix is defined as
\begin{align*}
    \Proj_{g_{ij}} = I_{d} - g_{ij}(t)g_{ij}^{T}(t),
\end{align*}
which is symmetric, positive semi-definite, and idempotent.

The formation \( (\mathcal{G}, q(t)) \) is defined as the combination of the interaction graph \( \mathcal{G} \) and the stacked position vector \( q(t) \). The definition of the target formation is introduced as follows.

\begin{Definition}[Target Formation]
The target formation \( (\mathcal{G}, q^{*}(t)) \) is a formation where the stacked  vector $$ q^{*}(t)=\textnormal{\mbox{col}}(q_{1}^{*}(t),\ldots, q_{n}^{*}(t)) $$ satisfies all  bearing constraints \( \{g_{ij}^{*}\}_{(i,j) \in \mathcal{E}} \) for all \( t \geq 0 \).
\end{Definition}
 
%Since \( g_{ij}(t) = -g_{ji}(t) \) and the underlying graph is undirected, it suffices to specify only half of the bearing constraints, which we label as \(  \{g_{ij}^{*}\}_{(i,j) \in \mathcal{E}} \). For notational convenience and without loss of clarity, the symbols \( g_{k}(t) \) and \( g_{ij}(t) \) are used interchangeably throughout this paper. The same applies to \( q_{\varepsilon_k}(t) \) and \( q_{ij}(t) \). The matrix $H \in \RR^{m \times n}$ is said to be incidence matrix, if $[H]_{ki}=1$ and $[H]_{kj}=-1$ if the $k$th edge $(i,j)\in \EE$, otherwise $0$.
 
  The Laplacian and bearing Laplacian matrices are defined as follows.

\begin{Definition}[Laplacian Matrix]
Given a graph \( \mathcal{G} \), the Laplacian matrix \( \mathcal{L} \in \mathbb{R}^{n \times n} \) is defined element-wise by
\begin{align*}
    \mathcal{L}_{ij} = \begin{cases}
        0, & \text{if } i \neq j \text{ and } (i,j) \notin \mathcal{E}; \\
        1, & \text{if } i \neq j \text{ and } (i,j) \in \mathcal{E}; \\
        {-\sum\limits_{i\in \NN_{i}} 1} , & \text{if } i = j.
    \end{cases}
\end{align*}
\end{Definition}

\begin{Definition}[Bearing Laplacian Matrix]
Given a graph \( \mathcal{G} \) and desired bearings \( \{g_{ij}^{*}\}_{(i,j) \in \mathcal{E}} \), the bearing Laplacian matrix \( \mathcal{B} \in \mathbb{R}^{nd \times nd} \) is defined block-wise by
\begin{align*}
    \mathcal{B}_{ij} = \begin{cases}
        \mathbf{0}, & \text{if } i \neq j \text{ and } (i,j) \notin \mathcal{E}; \\
        \Proj_{g_{ij}^{*}}, & \text{if } i \neq j \text{ and } (i,j) \in \mathcal{E}; \\
        -\sum\limits_{j \in \mathcal{N}_i} \Proj_{g_{ij}^{*}}, & \text{if } i = j.
    \end{cases}
\end{align*}
\end{Definition}

In line with the partition \( \mathcal{V} = \mathcal{V}_l \cup \mathcal{V}_f \), the Laplacian and bearing Laplacian matrices are partitioned as follows:
\begin{align*}
    \mathcal{L} = \begin{bmatrix}
        \mathcal{L}_{ll} & \mathcal{L}_{lf} \\
        \mathcal{L}_{fl} & \mathcal{L}_{ff}
    \end{bmatrix},
& &
    \mathcal{B} = \begin{bmatrix}
        \mathcal{B}_{ll} & \mathcal{B}_{lf} \\
        \mathcal{B}_{fl} & \mathcal{B}_{ff}
    \end{bmatrix},
\end{align*}
where
%\begin{align*}
	$\mathcal{L}_{ll} \in \mathbb{R}^{n_l \times n_l}$, $\mathcal{L}_{lf} \in \mathbb{R}^{n_l \times n_f},~\mathcal{L}_{fl} \in \mathbb{R}^{n_f \times n_l}$,~$\mathcal{L}_{ff}  \in \mathbb{R}^{n_f \times n_f}$,% \\
	~$\mathcal{B}_{ll} \in \mathbb{R}^{n_l d \times n_l d}$, $\mathcal{B}_{lf} \in \mathbb{R}^{n_l d \times n_f d}$, $\mathcal{B}_{fl} \in \mathbb{R}^{n_f d \times n_l d}$,%\\
	~{and} $\mathcal{B}_{ff} \in \mathbb{R}^{n_f d \times n_f d}$.
%\end{align*}

\begin{Remark}\label{uq_Rem}
Since the desired position vector \( q^{*}(t) \) satisfies the identity \( \mathcal{B}p^{*}(t) = \mathbf{0} \), it follows that
\begin{align}
    \mathcal{B}_{fl}q_{l}^{*}(t) + \mathcal{B}_{ff}q_{f}^{*}(t) = \mathbf{0}. \label{br_id}
\end{align}
Thus, given the desired leader positions \( q_{l}^{*}(t) \), if \( \mathcal{B}_{ff} \) is nonsingular, then the desired follower positions \( q_{f}^{*}(t) \) are uniquely determined by $$q_{f}^{*}(t) = -(\mathcal{B}_{ff})^{-1} \mathcal{B}_{fl}q_{l}^{*}(t).$$
\end{Remark}
 \begin{Remark}
 It can be verified that $\mathcal{L} \mathbf{1}_{n}=\mathbf{0}$ which implies the relation
   \begin{align}
 	\mathcal{L}_{fl} \mathbf{1}_{n_{l}}+\mathcal{L}_{ff}\mathbf{1}_{n_{f}}=\mathbf{0}. \label{Lp_eq}
 \end{align}
 \end{Remark}

   \subsection{Euler-Lagrange Systems}
  Following the formulations in \cite{li2020adaptive} and \cite{zhao2023adaptive}, for all $i \in \VV_f$, the dynamics of each Euler–Lagrange agent are described by
\begin{align}
    M_{i}(q_{i}) \ddot{q}_{i}(t) + C_{i}(q_{i}, \dot{q}_{i}) \dot{q}_{i}(t) + D_{i}(q_{i}) \dot{q}_{i}(t) = \tau_{i}(t), \label{fl_sys}
\end{align}
where \( M_{i}(q_{i}) \in \mathbb{R}^{d \times d} \) is the inertia matrix, \( C_{i}(q_{i}, \dot{q}_{i}) \in \mathbb{R}^{d \times d} \) is the Coriolis and centripetal matrix, \( D_{i}(q_{i}) \in \mathbb{R}^{d \times d} \) is the gravitational damping matrix, and \( \tau_{i}(t) \in \mathbb{R}^{d} \) denotes the control input (torque).

The Euler–Lagrange system possesses the following intrinsic properties:\\
\textit{Property 2.1:} The inertia matrix \( M_{i}(q_{i}) \) is symmetric and positive definite.\\
 \textit{Property 2.2:} For any auxiliary signal \( \zeta_{i}(t) \in \mathbb{R}^{d} \), the following identity holds:
    \begin{align}
        M_{i}(q_{i}) \dot{\zeta}_{i}(t) + C_{i}(q_{i}, \dot{q}_{i}) \zeta_{i}(t) &+ D_{i}(q_{i}) \dot{q}_{i}(t) \notag \\
        & =  Y_{i}(q_i, \dot{q}_i, \zeta_i, \dot{\zeta}_i) \theta_{i}, \label{id1}
    \end{align}
    where \( \theta_{i} \in \mathbb{R}^{r_{i}} \) is a constant parameter vector, and \(  Y_{i}(q_i, \dot{q}_i, \zeta_i, \dot{\zeta}_i) \in \mathbb{R}^{d \times r_{i}} \) is the known regressor matrix dependent on \( q_{i}, \dot{q}_{i}, \zeta_{i}, \) and \( \dot{\zeta}_{i} \).\\ 
 \textit{Property 2.3:} For all \( q_{i}(t), \dot{q}_{i}(t) \in \mathbb{R}^{d} \), the matrix \( \dot{M}_{i}(q_{i}) - 2C_{i}(q_{i}, \dot{q}_{i}) \) is skew-symmetric.
\begin{Remark}\label{EL_pro}
	By Property~2.1, there exists a scalar \( \underline{m}_i > 0 \) such that \(M_i(t) \geq \underline{m}_i I_d\), and  $ M_i^{-1}(t) \leq \underline{m}_i^{-1} I_d$.
\end{Remark}

For notational simplicity and without loss of clarity, the matrices 
\(M_{i}(q_{i}), C_{i}(q_{i},\dot{q}_{i}), D_{i}(q_{i})\), and  \( Y_i(q_i, \dot{q}_i, \zeta_i, \dot{\zeta}_i) \) will be denoted by \(M_{i} , C_{i} , D_{i} \) and \( Y_i \) respectively in the remainder of the paper.

\subsection{Problem Formulation}
 
  We consider {the problem of tracking formations driven by bearings}  for \( n \) agents, where the leaders move with a common velocity \( v_{c}(t) \in \mathbb{R}^{d} \) generated by the following linear (leader) system:
\begin{subequations}\label{ld_sys}
\begin{align}
    \dot{\eta}(t) &= S\eta(t), \\
    v_{c}(t) &= F\eta(t),
\end{align}
\end{subequations}
where \( \eta(t) \in \mathbb{R}^{w} \) is the state, \( S \in \mathbb{R}^{w \times w} \), and \( F \in \mathbb{R}^{d \times w} \). The followers are governed by the Euler–Lagrange model given in~\eqref{fl_sys}, for all $i \in \VV_f$.

The control law \( \tau_{i}(t) \) to be designed for each follower takes the following form: for all $i \in \VV_{f}$,
\begin{subequations}\label{law_form}
\begin{align}
    \tau_{i}(t) &= H_{i}(\mathbf{q}_{i}(t), \xi_{i}(t), \{\xi_{j}(t)\}_{j \in \mathcal{N}_{i}}), \\
    \dot{\xi}_{i}(t) &= G_{i}(\mathbf{q}_{i}(t), \xi_{i}(t), \{\xi_{j}(t)\}_{j \in \mathcal{N}_{i}}),
\end{align}
\end{subequations}
where $\mathbf{q}_{i}(t)=\mbox{col}(q_{i}(t), \dot{q}_{i}(t), \ddot{q}_{i}(t))$, \( \xi_{i}(t) \in \mathbb{R}^{s} \) is an auxiliary variable with dimension $s$ to be specified, and \( H_{i}(\cdot) \) and \( G_{i}(\cdot) \) are {global} functions to be determined. Since the control law~\eqref{law_form} depends only on the local information of agent \( i \) and its neighbors, it is said to be distributed.

We now formally state the problem of interest.

\begin{Problem}\label{Pro1}
Given an interaction graph \( \mathcal{G} \), a set of bearing constraints \(  \{g_{ij}^{*}\}_{(i,j) \in \mathcal{E}} \), the leader system~\eqref{ld_sys}, and followers governed by~\eqref{fl_sys}. Further  assume that $q_{a}(t)=q_{a}^{*}(t)$, for all $t \geq 0$. We seek to design a distributed control law of the form~\eqref{law_form} such that the followers' positions \( q_{f}(t) \) asymptotically converges to the desired positions \( q_{f}^{*}(t)\), i.e., mathematically
\begin{align*}
    \lim_{t \to \infty} \big( q_f(t) - q_f^{*}(t) \big) = \mathbf{0}.
    \end{align*}
\end{Problem}

The solvability of Problem~\ref{Pro1} is based on the following assumptions.

\begin{Assumption}\label{uq_amp}
The matrix \( \mathcal{B}_{ff} \) is nonsingular.
\end{Assumption}

\begin{Assumption}\label{col_amp}
No collisions occur among agents during the formation evolution. %{\color{red} Can this be stated as an assumption? {\color{blue}SMW: Assumption~\ref{col_amp}, which is commonly adopted in the literature~\cite{tang2021formation, trinh2021robust}, ensures the avoidance of inter-agent collisions during the formation process, thereby guaranteeing safety throughout the maneuver. From a mathematical standpoint, the bearing becomes undefined when two agents coincide. Hence, this assumption is necessary.}{\color{red}MG: All good!}}
\end{Assumption}

\begin{Remark}
 Assumption~\ref{uq_amp} is the bearing-based localization condition as stated in \cite{zhao2016localizability}. As detailed in Remark~\ref{uq_Rem}, this assumption ensures the uniqueness of the target formation. In particular, it guarantees that the desired follower positions are uniquely determined by the leader positions and the bearing constraints. Assumption~\ref{col_amp} is a standard assumption \citep*{tang2021formation,trinh2021robust} used to avoid inter-agent collisions during the formation process. As will be shown in the next section, this assumption can be relaxed by imposing an appropriate condition on the initial formation.% {\color{red} If it can be relaxed then why do we need to include it?{\color{blue}: Theorem 1 still need this assumption} ok. i agree}
\end{Remark}
 
\section{Main Results}\label{sec three}

In this section, we aim to design a distributed control law of the form~\eqref{law_form} to solve Problem~\ref{Pro1}. To this end, we first develop a distributed observer to estimate the state of the leader system. Based on this observer, we then synthesize the control law by incorporating an adaptive mechanism to address parametric uncertainties and achieve convergence to the target formation. Finally, we derive a sufficient condition, dependent on the initial formation configuration, to ensure collision avoidance throughout the formation evolution.

\subsection{Distributed Observer}

 To estimate the state of the leader system \eqref{ld_sys}, we further need to following assumption.
 \begin{Assumption}\label{dec_amp}
 	The pair $(S, F)$ is detectable.
 \end{Assumption}

According to \cite[Theorem 3]{kucera1972contribution}, under Assumption \ref{dec_amp}, there exists a positive definite symmetric matrix $P \in \RR^{w \times w}$ such that 
\begin{align}
	PS^{T}+SP-PF^{T}FP+I_{w}=\mathbf{0}. \label{Ric_eq}
\end{align}
Let $\hat{v}_{i}(t) \in \RR^{d}$ be the estimate of $  v_{c}(t)$, for all $i \in \VV$ and, for all $i \in \VV_l$ and $t \geq 0$, let $\hat{v}_{i}(t)= v_{c}(t)$. The proposed dynamic compensator for system \eqref{ld_sys}, $\forall i \in \VV_{f}$, is given as follows:
\begin{subequations}\label{dis_ob1}
\begin{align}
	\dot{\hat{\eta}}_{i}(t)=&\ S{\hat{\eta}}_{i}(t)-\gamma L\sum_{j \in \NN_{i}}{\hat{v}}_{ij}(t)\\
	\hat{v}_{i}(t)=&\ F{\hat{\eta}}_{i}(t) \label{dis_ob1_b}
\end{align}	
\end{subequations}
where $L= PF^{T}$, $\gamma>0$ is to be determined, and $${\hat{v}}_{ij}(t)={\hat{v}}_{i}(t)-{\hat{v}}_{j}(t).$$
If the dynamic compensator \eqref{dis_ob1} is such that $$\lim_{t \to \infty}({\hat{\eta}}_{i}(t)-\eta(t))=\mathbf{0},$$ then we call it a distributed observer for the leader system \eqref{ld_sys}.

To compactly  express the system, define the following notations 
\begin{align}
	{\hat{\eta}}_{f}=&\ \textnormal{\mbox{col}}({\hat{\eta}}_{n_{l}+1},\ldots,{\hat{\eta}}_{n}),\quad {\eta}_{f}^{*}=\mathbf{1}_{n_{f}}\otimes \eta, \quad {\tilde{\eta}}_{f}= {\hat{\eta}}_{f}-{\eta}_{f}^{*},\notag \\
	\hat{v}_{f}=&\ \textnormal{\mbox{col}}(\hat{v}_{n_{l}+1},\ldots,\hat{v}_{n}), \quad v_{f}^{*}=\mathbf{1}_{n_{f}}\otimes v_{c},
	 \quad \tilde{v}_{f} = v_{f}-v_{f}^{*}. \notag %\label{not1}
\end{align}
Then, we are able to rewrite \eqref{dis_ob1} into the following compact form:
\begin{subequations}\label{dis_ob2}
\begin{align}
	\dot{\tilde{\eta}}_{f}(t)=&\ (I_{n_{f}}\otimes S){\tilde{\eta}}_{f}(t)-f(\eta(t), \hat{\eta}_f(t))\label{dis_ob2_a}\\
	\tilde{v}_{f} (t)=&\ (I_{n_{f}}\otimes F){\tilde{\eta}}_{i}(t) \label{dis_ob2_b}
	\end{align}	
\end{subequations}
where \[f(\eta(t), \eta_f(t))=\gamma (\mathcal{L}_{fl}\otimes LF)(\mathbf{1}_{n_l}\otimes {\eta}(t))+\gamma (\mathcal{L}_{ff}\otimes LF)\hat{\eta}_{f}(t).\]
Using the identity \eqref{Lp_eq} converts \eqref{dis_ob2_a} into 
\begin{align}\label{dis_ob3}
	\dot{\tilde{\eta}}_{f}(t)=&\ S_f {\tilde{\eta}}_{f}(t)
\end{align}
where  $S_f \triangleq (I_{n_f} \otimes S) - \gamma (\mathcal{L}_{ff} \otimes L F)$.

To analyze the convergence of the system \eqref{dis_ob3}, we need the following fundamental lemma.
\begin{Lemma}\label{lem1}
	Under Assumption \ref{uq_amp}, for each follower, there exists at least one leader that has a directed path to it.
\end{Lemma}
\begin{Proof}
	 We will prove the statement by contradiction. Suppose that there exists at least one follower that is not connected to any leader through any path in the graph. Without loss of generality, we label such followers as \( \{n_{f_1}, \ldots, n\} \), for some \( n_{f_1} \in \mathcal{V}_{f} \). Under this assumption, the bearing Laplacian matrix \( \mathcal{B} \) and the submatrix \( \mathcal{B}_{ff} \) can be partitioned as follows:
\begin{align*}
\mathcal{B} = \left[
\begin{array}{@{}c:c@{}}
\mathcal{B}_{ll} & \begin{matrix}
[\mathcal{B}_{lf}]_{1} & \mathbf{0}
\end{matrix} \\
\hdashline
\begin{matrix}
[\mathcal{B}_{fl}]_{1} \\
\mathbf{0}
\end{matrix} &
\begin{matrix}
[\mathcal{B}_{ff}]_{11} & \mathbf{0} \\
\mathbf{0} & [\mathcal{B}_{ff}]_{22}
\end{matrix}
\end{array}
\right],  
\mathcal{B}_{ff} = \begin{bmatrix}
[\mathcal{B}_{ff}]_{11} & \mathbf{0} \\
\mathbf{0} & [\mathcal{B}_{ff}]_{22}
\end{bmatrix}
\end{align*}
where 
$[\mathcal{B}_{lf}]_{1} \in \mathbb{R}^{n_{l}d \times \omega_1 d}$,
	$[\mathcal{B}_{fl}]_{1} \in  \mathbb{R}^{\omega_1 d \times n_{l}d}$,
	$[\mathcal{B}_{ff}]_{11} \in \mathbb{R}^{\omega_1 d \times \omega_1 d}$ and
	$[\mathcal{B}_{ff}]_{22} \in \mathbb{R}^{\omega_2 d \times \omega_2 d}$
%\end{align*}
 are non-zero matrices with $\omega_1=n_{f_1}-n_{l} $ and $\omega_2=n-n_{f_1} $.

Let \( x = \mathrm{col}(\mathbf{0}, \mathbf{1}_{\omega_2 d}) \in \mathbb{R}^{n_{f}d} \). Then a direct computation yields:
\begin{align*}
\mathcal{B}_{ff}x = \begin{bmatrix}
\mathbf{0} \\
[\mathcal{B}_{ff}]_{22} \mathbf{1}_{\omega_2 d}
\end{bmatrix} = \mathbf{0}.
\end{align*}

This implies that \( x \) lies in the null space of \( \mathcal{B}_{ff} \), and hence \( \mathcal{B}_{ff} \) is singular, which contradicts Assumption~\ref{uq_amp}. Therefore, every follower must be connected to at least one leader via a path in the graph. The proof is complete.
\end{Proof}

\begin{Lemma}\citep*[Lemma 5.1]{ren2010distributed}\label{lem2}
	If each follower has a directed path to at least one leader, then all eigenvalues of $L_{ff}$ have positive real parts.
\end{Lemma}

We have the following result about the convergence of \eqref{dis_ob3}.
 \begin{Lemma}\label{dis_ob}
Under Assumptions~\ref{uq_amp} and~\ref{dec_amp}, if $$ 2\lambda_{\min}(\mathcal{L}_{ff})\gamma > 1 ,$$ then system~\eqref{dis_ob3} is such that \( \lim_{t \to \infty} \tilde{\eta}_{f}(t) = \mathbf{0} \) exponentially. Consequently, \( \lim_{t \to \infty} \tilde{v}_{f}(t) = \mathbf{0} \) exponentially.
\end{Lemma}

\begin{Proof}
By Lemma~\ref{lem1} and under Assumption~\ref{uq_amp}, each follower is connected to at least one leader. Therefore, by Lemma \ref{lem2}, the submatrix \( \mathcal{L}_{ff} \) is positive definite.

Since the graph \( \mathcal{G} \) is undirected, \( \mathcal{L}_{ff} \) is symmetric. Hence, it admits an eigen-decomposition:
\[
\mathcal{L}_{ff} = T \Lambda_{ff} T^{\top},
\]
where \( T \in \RR^{n_f \times n_f }\) is an orthogonal matrix and $$ \Lambda_{ff} = \textnormal{diag}(\lambda_1(\mathcal{L}_{ff}) , \ldots, \lambda_{n_f}(\mathcal{L}_{ff}). $$
Then, the eigenvalues of \( S_f \) coincide with those of the matrices
\begin{align*}
S - \gamma\lambda_{k}(\mathcal{L}_{ff}) L F, & &\forall k =1,\ldots, n_f  .
\end{align*}
Consider
{
\begin{align*}
(S - \gamma \lambda_{k}(\mathcal{L}_{ff}) L F) P& + P (S - \gamma \lambda_{k}(\mathcal{L}_{ff}) L F)^{\top} \\
&= S P + P S^{\top} - 2 \gamma \lambda_{k}(\mathcal{L}_{ff}) P F^{\top} F P \\
&\leq S P + P S^{\top} - 2 \gamma \lambda_{\min}(\mathcal{L}_{ff})  P F^{\top} F P\\
&\leq -I_w
\end{align*}}
where the last inequality is due to  \eqref{Ric_eq} and the condition $2\lambda_{\min}(\mathcal{L}_{ff})\gamma > 1$. This implies that each matrix \( S - \gamma \lambda_k(\mathcal{L}_{ff}) L F \), for all $k=1, \ldots, n_f$, is Hurwitz, and thus the matrix \( S_f \) is also Hurwitz. Consequently, \( \tilde{\eta}_{f}(t) \to \mathbf{0} \) exponentially, and hence \( \tilde{v}_{f}(t) \to \mathbf{0} \) exponentially as well. This completes the proof.
\end{Proof}

\begin{Remark}
 Compared with the conventional distributed observer presented in~\cite{Huang2022}, the proposed design exhibits at least two notable advantages. First, the convergence of the distributed observer in~\cite{Huang2022} relies on the connectivity condition of the underlying graph, whereas the proposed approach is built upon the bearing-based localization condition. This condition not only ensures the uniqueness of the target formation but also guarantees the exponential convergence of the dynamic compensator. Second, the design accommodates the presence of multiple leaders, while the conventional method is limited to handling the single-leader scenario.
\end{Remark}

\subsection{Formation Tracking Control}

With the aid of the distributed observer given in \eqref{dis_ob2}, we now proceed to design a distributed control law for the Euler–Lagrange agents governed by the model \eqref{fl_sys}. To facilitate the control design, we first introduce the following variables:
\begin{align}
    \zeta_{i}(t) &= F  \hat{\eta}_{i}(t) + \sum_{j \in \mathcal{N}_{i}} \Proj_{g_{ij}^{*}}q_{ij}(t)\notag \\
    \dot{\zeta}_{i}(t) &= F \dot{\hat{\eta}}_{i}(t) + \sum_{j \in \mathcal{N}_{i}} \Proj_{g_{ij}^{*}}\dot{q}_{ij}(t) \notag \\
    s_{i}(t) &= \dot{q}_{i}(t) - \zeta_{i}(t). \notag
\end{align}

We now present the distributed control law:
{
\begin{subequations}\label{law1}
\begin{align}
    \tau_{i}(t) &= Y_{i}  \hat{\theta}_{i}(t) - \Lambda_{s_{i}} s_{i}(t) \\
    \dot{\hat{\theta}}_{i}(t) &= -\Lambda_{\theta_{i}}  Y_{i}^{\top} s_{i}(t) \\
    \dot{\hat{\eta}}_{i}(t) &= S \hat{\eta}_{i}(t) - \gamma L \sum_{j \in \mathcal{N}_{i}} \hat{v}_{ij}(t)
\end{align}
\end{subequations}}
where \(Y_{i}  \) is the regressor matrix defined in \eqref{id1}, and \( \Lambda_{s_{i}} \in \RR^{d \times d}\), and  \( \Lambda_{\theta_{i}} \in \RR^{r_{i} \times r_i} \) are some positive definite gain matrices.

Substituting the control law \eqref{law1} into the agent dynamics \eqref{fl_sys}, the closed-loop system is given by:
{
\begin{subequations}\label{close1}
\begin{align}
    M_{i} \ddot{q}_{i}(t) + C_{i} \dot{q}_{i}(t) + D_{i} \dot{q}_{i}(t) &=  Y_{i} \hat{\theta}_{i}(t) - \Lambda_{s_{i}} s_{i}(t) \label{close1_a} \\
    \dot{\hat{\theta}}_{i}(t) &= -\Lambda_{\theta_{i}}  Y_{i}^{\top} s_{i}(t) \label{close1_b} \\
    \dot{\hat{\eta}}_{i}(t) &= S \hat{\eta}_{i}(t) - \gamma L \sum_{j \in \mathcal{N}_{i}} \hat{v}_{ij}(t).
\end{align}
\end{subequations}}
It follows from Property 2.2 that one has
\begin{align*}
    M_{i}  \dot{\zeta}_{i}(t) + C_{i}  \zeta_{i}(t) + D_{i}  \dot{q}_{i}(t)
    =  Y_{i} \theta_{i}.
\end{align*}
Using this relation, the closed-loop dynamics given in~\eqref{close1} can be reformulated as:
\begin{subequations} \label{close2}
\begin{align}
    \dot{s}_{i}(t) =& -M_{i}^{-1}  ( C_{i}  s_{i}(t)  - Y_{i} \tilde{\theta}_{i}(t)+ \Lambda_{s_{i}} s_{i}(t) ) \\
    \dot{\tilde{\theta}}_{i}(t) =& -\Lambda_{\theta_{i}}  Y_{i}^{T} s_{i}(t) \\
    \dot{\hat{\eta}}_{i}(t) =& S \hat{\eta}_{i}(t) - \gamma L \sum_{j \in \mathcal{N}_{i}} \hat{v}_{ij}(t) 
\end{align}
\end{subequations} where \( \tilde{\theta}_{i}(t) = \hat{\theta}_{i}(t) - \theta_{i} \).\\

To represent the compact form of the closed-loop system, we define the following notation: $C_{f} = \text{\textnormal{blkdiag}}(C_{n_l+1}, \ldots, C_{n} )$,
\begin{align}
    M_{f}  &= \text{\textnormal{blkdiag}}(M_{n_l+1} , \ldots, M_{n} ), 
    Y_{f} = \text{\textnormal{blkdiag}}(Y_{n_l+1}, \ldots, Y_{n}) \notag \\
    \Lambda_{\theta_f} &= \text{\textnormal{blkdiag}}(\Lambda_{\theta_{n_l+1}}, \ldots, \Lambda_{\theta_{n}}), \tilde{\theta}_{f} = \text{col}(\tilde{\theta}_{n_l+1}, \ldots, \tilde{\theta}_{n}), \notag \\
    \Lambda_{s_f} &= \text{\textnormal{blkdiag}}(\Lambda_{s_{n_l+1}}, \ldots, \Lambda_{s_{n}}),
    s_{f}= \text{col}(s_{n_l+1}, \ldots, s_{n}). \label{not_sf} %\\
    %.\notag 
\end{align}
With these notations, we obtain the compact form of the closed-loop system \eqref{close2}:
\begin{subequations}\label{close3}
\begin{align}
    \dot{s}_{f}(t) &= -M_{f}^{-1}(  C_{f} s_{f}(t)- Y_{f} \tilde{\theta}_{f}(t)  +\Lambda_{s_f} s_{f}(t) ) \label{close3_a} \\
    \dot{\tilde{\theta}}_{f}(t) &= -\Lambda_{\theta_f} Y_{f}^{T}(t) s_{f}(t) \\
    \dot{\tilde{\eta}}_{f}(t) &= S_{f} \tilde{\eta}_{f}(t).  \label{close3_c}
\end{align}
\end{subequations}

\begin{Lemma}\citep*{Huang2022}\label{peb_lem}
	Consider the following linear system:
	\begin{align}
	\dot{x}(t)=Ax(t)+F(t)	\label{lem_peb_eq1}
	\end{align}
 where $x(t) \in \RR^{s}$, $A \in \RR^{s \times s}$ is Hurwitz, and $F(t) \in \RR^{s \times s}$ is piecewise continuous and uniformly continuous over $[0,\infty)$. Then, for any $x(0)$, the solution of \eqref{lem_peb_eq1} tends to zero asympotically if $F(t) \rightarrow \mathbf{0}$ as $t \rightarrow \infty$.
\end{Lemma}

To establish the stability properties of the closed-loop system \eqref{close3}, like \cite[Section 5]{Huang2022}, we impose the following two additional standard assumptions.

\begin{Assumption}\label{ld_amp}
The matrix \( S \) has no eigenvalues with positive real parts, and the signal \( \eta(t) \) remains uniformly bounded for all \( t \in [0, \infty) \).
\end{Assumption}

\begin{Assumption}\label{para_amp}
If \( \dot{q}_{i}(t) \), \( \zeta_{i}(t) \), and \( \dot{\zeta}_{i}(t) \) are uniformly bounded over \( [0, \infty) \), then both \( C_{i}(q_i(t), \dot{q}_i(t)) \) and \( Y_{i}(q_i(t), \dot{q}_i(t), \zeta_{i}(t), \dot{\zeta}_{i}(t)) \) are also uniformly bounded over \( [0, \infty) \), for all \( i \in \mathcal{V}_f \).
\end{Assumption}

We are now ready to present the main result on the convergence of the closed-loop system.

\begin{Theorem}\label{Thm1}
Suppose that Assumptions~\ref{uq_amp} to \ref{para_amp} hold. If the control gain \( \gamma \) satisfies $$2\lambda_{\min}(\mathcal{L}_{ff})\gamma > 1,$$ then the control law \eqref{law1} ensures that $$\lim_{t \to \infty} \big( q(t) - q^{*}(t) \big) = \mathbf{0},$$ and hence, Problem~\ref{Pro1} is solved.
\end{Theorem}

\begin{Proof}
	We begin by considering the following Lyapunov function candidate:
\begin{align}
    V_{1}(s_{f}(t), \tilde{\theta}_{f}(t)) = s_{f}^{T}(t) M_f s_{f}(t) + \tilde{\theta}_{f}^{T}(t) \Lambda_{\theta_{f}}^{-1} \tilde{\theta}_{f}(t). \label{ly1}
\end{align}
whose time derivative  along the trajectories of the closed-loop system \eqref{close3} is given by
\begin{align}
    \dot{V}_{1}(s_{f}(t), \tilde{\theta}_{f}(t)) 
    =&\ 2 s_{f}^{T}(t) M_f  \dot{s}_{f}(t) + s_{f}^{T}(t) \dot{M}_f  s_{f}(t) \notag \\
    &+ 2 \tilde{\theta}_{f}^{T}(t) \Lambda_{\theta_{f}}^{-1} \dot{\tilde{\theta}}_{f}(t) \notag \\
    =&\ 2 s_{f}^{T}(t) \left( - C_{f} s_{f}(t) + Y_{f}   \tilde{\theta}_{f}(t) - \Lambda_{s_f} s_{f}(t) \right) \notag \\
    & + s_{f}^{T}(t) \dot{M}_f s_{f}(t) - 2 \tilde{\theta}_{f}^{T}  Y_{f}^{T}  s_{f}(t) \notag \\
    =&\ s_{f}^{T}(t) \left( \dot{M}_f  - 2 C_{f} \right) s_{f}(t) - 2 s_{f}^{T}(t) \Lambda_{s_f} s_{f}(t). %\label{d_ly1} %\notag 
\end{align}
According to Property 2.3, for {each} \( i \in \mathcal{V}_f \), the matrix \( \dot{M}_{i}  - 2 C_{i}  \) is skew-symmetric. Consequently, the block-diagonal matrix \( \dot{M}_f  - 2 C_f  \) is also skew-symmetric, and hence,
\begin{align}
    \dot{V}_{1}(s_{f}(t), \tilde{\theta}_{f}(t)) = \underbrace{-2 s_{f}^{T}(t) \Lambda_{s_f} s_{f}(t)}_{ \leq 0}. \label{d_ly1_2}
\end{align}
Then, taking the second derivative yields
\begin{align}
    \ddot{V}_{1}(s_{f}(t), \tilde{\theta}_{f}(t)) = -2 s_{f}^{T}(t) \Lambda_{s_f} \dot{s}_{f}(t). \label{dd_ly1}
\end{align}
From~\eqref{d_ly1_2}, it follows that the function \( V_1(s_f(t), \tilde{\theta}_f(t)) \) is non-increasing and uniformly bounded. Consequently, both \( s_f(t) \) and \( \tilde{\theta}_f(t) \) remain uniformly bounded for all \( t \geq 0 \). To further establish the boundedness of the second derivative \( \ddot{V}_1(s_f(t), \tilde{\theta}_f(t)) \) in~\eqref{dd_ly1}, it suffices to demonstrate that the derivative \( \dot{s}_f(t) \) is also uniformly bounded. To this end, using the relation \eqref{dis_ob1_b}, we now express the variable \( s_f(t) \) in the following compact form:
\begin{align}
    s_{f}(t) = \dot{q}_{f}(t) - \hat{v}_{f}(t) + \mathcal{B}_{fa} q_{a}(t) + \mathcal{B}_{ff} q_{f}(t), \label{thm_eq1}
\end{align}
which, together with the identity in \eqref{br_id}, leads to the error dynamics:
\begin{align}
    \dot{\tilde{q}}_{f}(t) = -\mathcal{B}_{ff} \tilde{q}_{f}(t) + d_f(t), \label{thm_sys1}
\end{align}
where \( \tilde{q}_{f}(t) = q_{f}(t) - q_{f}^{*}(t) \), and \( d_f(t) = s_{f}(t) +  \tilde{v}_{f}(t) \), with \(  \tilde{v}_{f}(t) \) defined in \eqref{dis_ob2_b}.

By Lemma~\ref{dis_ob}, the observer output error \(  \tilde{v}_{f}(t) \) is uniformly bounded. From \eqref{d_ly1_2}, the signal \( s_{f}(t) \) is also uniformly bounded, implying that \( d_f(t) \) is uniformly bounded as well. That is, there exists a constant \( \bar{D} > 0 \) such that
\begin{align*}
\|d_f(t)\| \leq \bar{D}, & & \forall t \geq 0.
\end{align*}
The solution of \eqref{thm_sys1} can be expressed as follows:
\begin{align*}
    \tilde{q}_{f}(t) &= \exp(-\mathcal{B}_{ff} t) \tilde{q}_{f}(0) + \int_{0}^{t} \exp(-\mathcal{B}_{ff}(t - \tau)) d_f(\tau) \, d\tau,
\end{align*}
from which we deduce
\begin{align}
    \|\tilde{q}_{f}(t)\| \leq&\ \|\exp(-\mathcal{B}_{ff} t)\| \cdot \|\tilde{q}_{f}(0)\| \notag \\
    &+ \bar{D} \int_{0}^{t} \|\exp(-\mathcal{B}_{ff}(t - \tau))\| d\tau \notag  \\
    \leq& \exp(-\lambda_{\min}(\mathcal{B}_{ff}) t) \|\tilde{q}_{f}(0)\| + \tfrac{\bar{D}}{\lambda_{\min}(\mathcal{B}_{ff})}. \label{thm_eq2}
\end{align}
This implies that \( \tilde{q}_{f}(t) \) is uniformly bounded for all \( t \geq 0 \). Moreover, uniform boundedness of \( \tilde{q}_{f}(t) \) and \eqref{thm_sys1} implies that \( \dot{\tilde{q}}_{f}(t) \) is also uniformly bounded.

Next, by differentiating the identity \eqref{br_id} and using the relation \( \dot{q}_{l}^{*}(t) = \mathbf{1}_{n_l} \otimes v_{c}(t) \), it follows that $$ \dot{q}_{f}^{*}(t) = \mathbf{1}_{n_f} \otimes v_{c}(t), $$ which is uniformly bounded under Assumption~\ref{ld_amp}. Consequently, \( \dot{q}_{f}(t) \) is uniformly bounded.

Additionally, the uniform boundedness of \( \tilde{q}_{f}(t) \) and \( \dot{\tilde{q}}_{f}(t) \) implies that \( \mathcal{B}_{ff}\tilde{q}_{f}(t) \) and \( \mathcal{B}_{ff}\dot{\tilde{q}}_{f}(t) \)  are uniformly bounded whose block-entries are  \( \sum\limits_{j \in \mathcal{N}_{i}} \Proj_{g_{ij}^{*}} q_{ij}(t) \) and \( \sum\limits_{j \in \mathcal{N}_{i}} \Proj_{g_{ij}^{*}} \dot{q}_{ij}(t) \) respectively, for all \( i \in \mathcal{V}_f \).

From Lemma~\ref{dis_ob}, \( \tilde{\eta}_f(t) \) is uniformly bounded. Combined with Assumption~\ref{ld_amp}, this implies that \( \hat{\eta}_{i}(t) \) is uniformly bounded for all \( i \in \mathcal{V}_f \). Then, by Assumption~\ref{para_amp}, both \(  C_{f} \) and \( Y_{f} \) are uniformly bounded. By Remark \ref{EL_pro}, \( M_f^{-1}  \) is upper bounded. Since \( s_f(t) \) is uniformly bounded, in view of \eqref{close3_a}, it follows that \( \dot{s}_f(t) \) is uniformly bounded.

Consequently, from \eqref{dd_ly1}, \( \ddot{V}(t) \) is bounded. Applying Barbalat’s Lemma \citep*[Lemma 4.2]{Slotine91} to the function \eqref{ly1} yields
$$\lim_{t \to \infty} \dot{V}_{1}(s_{f}(t), \tilde{\theta}_{f}(t)) = 0,$$
which, together with \eqref{d_ly1_2},  implies \(\lim_{t \to \infty} s_{f}(t) = \mathbf{0}\).

Finally, by Lemma~\ref{dis_ob}, we also have \( \tilde{v}_{f}(t) \to \mathbf{0} \) exponentially. Therefore, the perturbed linear system \eqref{thm_sys1} satisfies the standard input-to-state stability (ISS) form of \eqref{lem_peb_eq1} with \( x(t)=\tilde{q}_f(t) \), \( A = -\mathcal{B}_{ff} \) being Hurwitz, and \(F(t)= d_f(t) \). Hence, it follows that
$$\lim_{t \to \infty} \tilde{q}_{f}(t) = \mathbf{0},$$
which completes the proof.
\end{Proof}

\subsection{Collision Avoidance}

In this section, we establish a sufficient condition for ensuring inter-agent collision avoidance throughout the formation process based on the initial formation.

From the proof of Lemma~\ref{dis_ob}, the matrix \( S_f \) is Hurwitz. Therefore, there exists a symmetric positive definite matrix { \( P_{f} \in \mathbb{R}^{\omega n_f \times \omega n_f} \) such that
\begin{align*}
    S_f^{T} P_{f} + P_f S_f = -I_{\omega n_f}.
\end{align*}}
Then, consider the function 
\begin{align}
	V_{2}(\tilde{\eta}_f(t))&= \tilde{\eta}_f^{T}(t)P_f \tilde{\eta}_f(t) \label{ly2}
\end{align}
whose time derivative along the trajectory of \eqref{close3_c} is
\begin{align}
	\dot{V}_{2}(\tilde{\eta}_f(t))=&\underbrace{-\|\tilde{\eta}_f(t)\|^2}_{\leq 0}. \label{d_ly2}
\end{align}

We now present the following result.

\begin{Proposition} \label{Prop1}
Under Assumptions~\ref{uq_amp}, \ref{dec_amp}, \ref{ld_amp}, and \ref{para_amp}, if the initial condition of the closed-loop system \eqref{close3} satisfies
\begin{align}
    &\inf_{t \geq 0,~i,j \in \mathcal{V},~i \neq j} \|q_{i}^{*}(t) - q_{j}^{*}(t)\|\geq  \notag \\
    &\quad \sqrt{n} \left( \|\tilde{q}_{f}(0)\| + \tfrac{\bar{D}}{\lambda_{\min}(\mathcal{B}_{ff})} \right) + \gamma, \label{col_cri}
\end{align}
where \( \bar{D} \) is defined as \(\bar{D} = D_{s_f} + D_{ \tilde{v}_{f}}\) with
\begin{align*}
    D_{s_f} &= \sqrt{ \tfrac{ s_{f}^{T}(0) M_f(q_f(0)) s_{f}(0) + \tilde{\theta}_{f}^{T}(0) \Lambda_{\theta_f}^{-1} \tilde{\theta}_{f}(0) }{\min_{i \in \mathcal{V}_f} \{ \underline{m}_{i} \}}  }, \\
    D_{ \tilde{v}_{f}} &= \|F\| \cdot \sqrt{ \tfrac{1}{\lambda_{\min}(P_f)} \tilde{\eta}_{f}^{T}(0) P_f \tilde{\eta}_{f}(0) },
\end{align*}
then the motions of agents maintain a minimum safety distance \( \gamma \) for all \( t \geq 0 \), i.e.,
\[
\inf_{t \geq 0, \forall i,j \in \mathcal{V}, i \neq j} \|q_{i}(t) - q_{j}(t)\| \geq \gamma.
\]
\end{Proposition}

\begin{Proof}
We begin by estimating the bound on the observer output error \(  \tilde{v}_{f}(t) \). From the definition and the Lyapunov function candidate \( V_2(\tilde{\eta}_f(t)) \) defined in \eqref{ly2}, it follows that
\begin{align}
    \underbrace{\| \tilde{v}_{f}(t)\| }_{\| F \tilde{\eta}_{f}(t) \|} &\leq \|F\| \cdot \| \tilde{\eta}_{f}(t) \| \notag \\
    &\leq \|F\| \cdot \sqrt{ \tfrac{V_2(\tilde{\eta}_{f}(t))}{\lambda_{\min}(P_f)} } \notag \\
    &\leq \underbrace{\|F\| \cdot \sqrt{ \tfrac{V_2(\tilde{\eta}_{f}(0))}{\lambda_{\min}(P_f)} }}_{
    D_{ \tilde{v}_{f}}} , \label{pp1_eq1}
\end{align}
where the last inequality follows from the nonincreasing monotonicity of \( V_2(t)\) established via \( \dot{V}_2(t) \leq 0 \) in \eqref{d_ly2}.

Next, we estimate the bound on \( s_f(t) \) using the positive definiteness of \( M_f(q_f(t)) \). By Remark~\ref{EL_pro}, we have
\begin{align*}
    \|s_f(t)\| &\leq \sqrt{ \tfrac{ s_f^T(t) M_f(q_f(t)) s_f(t) }{ \inf_{t \geq 0} \{\lambda_{\min}(M_f(q_f(t)))\} } }\\
    &= \sqrt{ \tfrac{ s_f^T(t) M_f(t) s_f(t) }{ \min_{i \in \mathcal{V}_f} \{ \underline{m}_{i} \} } } \\
    &\leq \sqrt{ \tfrac{ V_1(s_f(t), \tilde{\theta}_f(t)) }{ \min_{i \in \mathcal{V}_f} \{ \underline{m}_{i} \} } }.
\end{align*}
Since \( V_1 (s_{f}(t), \tilde{\theta}_{f}(t))\) is non-increasing as shown in \eqref{d_ly1_2}, we further obtain
\begin{align}
    \|s_f(t)\| \leq \underbrace{\sqrt{ \tfrac{ V_1(s_f(0), \tilde{\theta}_f(0)) }{ \min_{i \in \mathcal{V}_f} \{ \underline{m}_{i} \} } }}_{D_{s_f}}. \label{pp1_eq2}
\end{align}

Combining \eqref{pp1_eq1} and \eqref{pp1_eq2}, we can now bound the disturbance term \( d_f(t) = s_f(t) +  \tilde{v}_{f}(t) \) in the perturbed dynamics \eqref{thm_sys1}:
\begin{align*}
    \|d_f(t)\| &\leq  \|s_f(t)\| + \| \tilde{v}_{f}(t)\| \\
    & \leq \underbrace{D_{s_f} + D_{ \tilde{v}_{f}} }_{ \bar{D}}.
\end{align*}

Now, from the bound on $d_f(t)$ in the perturbed linear system \eqref{thm_sys1}, and using the relation \eqref{thm_eq2}, we have
\begin{align}
    \|\tilde{q}_f(t)\| \leq \|\tilde{q}_f(0)\| + \tfrac{\bar{D}}{\lambda_{\min}(\mathcal{B}_{ff})}. \label{pp1_eq3}
\end{align}
To derive the inter-agent distance bound, we consider
\begin{align}
     \|q_i(t) - q_j(t)\| 
     =&\ \|q_i(t) - q_i^*(t)  \notag \\
     &+ q_i^*(t)- q_j^*(t) + q_j^*(t) - q_j(t)\| \notag \\
    \geq&\ \|q_i^*(t) - q_j^*(t)\| - \|q_i(t) - q_i^*(t)\|\notag \\
   & - \|q_j(t) - q_j^*(t)\| \notag \\
    \geq& \inf_{t \geq 0, \forall i ,j \in \VV ,  i\neq j} \|q_i^*(t) - q_j^*(t)\| - \sqrt{n} \|\tilde{q}_f(t)\|. \label{pp1_eq4}
\end{align}
Substituting the bound \eqref{pp1_eq3} and the condition \eqref{col_cri} into \eqref{pp1_eq4}, we conclude:
\begin{align*}
   \|q_i(t) - q_j(t)\| 
    \geq& \inf_{t \geq 0, \forall i ,j \in \VV,  i\neq j} \|q_i^*(t) - q_j^*(t)\| \\
    &- \sqrt{n} \left( \|\tilde{q}_f(0)\| + \tfrac{\bar{D}}{\lambda_{\min}(\mathcal{B}_{ff})} \right) 
    \geq \gamma.
\end{align*}

Therefore, all agents maintain a minimum distance of at least \( \gamma \) for all \( t \geq 0 \), completing the proof.
\end{Proof}

% {\color{red} Should there be a statement/remark concerning the conservative nature of these bounds? how does it compare with other works?}
% \textcolor{blue}{Haoshu: Thanks for this advice. I  have added a remark including comparison and the conservative nature. Please see below. }
% \textcolor{blue}{
\begin{Remark}
  Proposition~\ref{Prop1} establishes a sufficient condition based on the initial values of the closed-loop system. Similar to the existing results in~\cite{zhao2015bearing} and \cite{trinh2018bearing}, the safe distance is ensured by imposing constraints on the initial conditions of the closed-loop system. When the formation is sufficiently close to the target configuration, \eqref{col_cri} guarantees that the minimal inter-agent distance is maintained for a small~$\gamma$. However, due to the higher complexity and nonlinearity of Euler--Lagrange systems, the existing methods in~\cite{zhao2015bearing} and \cite{trinh2018bearing} are applicable only to simple dynamics, such as single- or double-integrator models. The novelty of this work lies in the development of a new control design for Euler--Lagrange systems, which is able to ensure inter-agent collision avoidance, a capability not addressed in related studies such as~\cite{zhao2023adaptive} and \cite{li2020adaptive}.%{\color{red} Excellent!}
\end{Remark}
%}
\section{An Numerical Example}\label{sec four}
     \begin{figure}%[H]
    \centering
    \includegraphics[width = 0.25\textwidth]{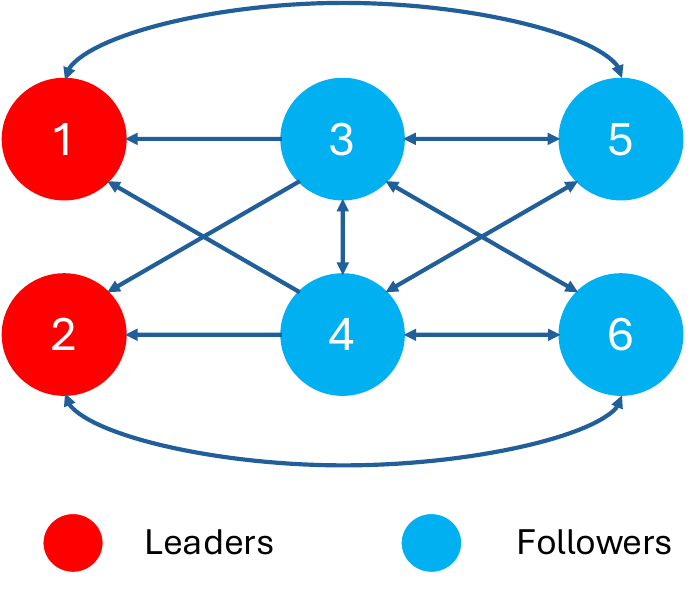}
    \caption{Communication Graph}
    \label{fig:graph}
\end{figure}

\begin{figure}%[H]
    \centering
    \includegraphics[width = 0.5\textwidth]{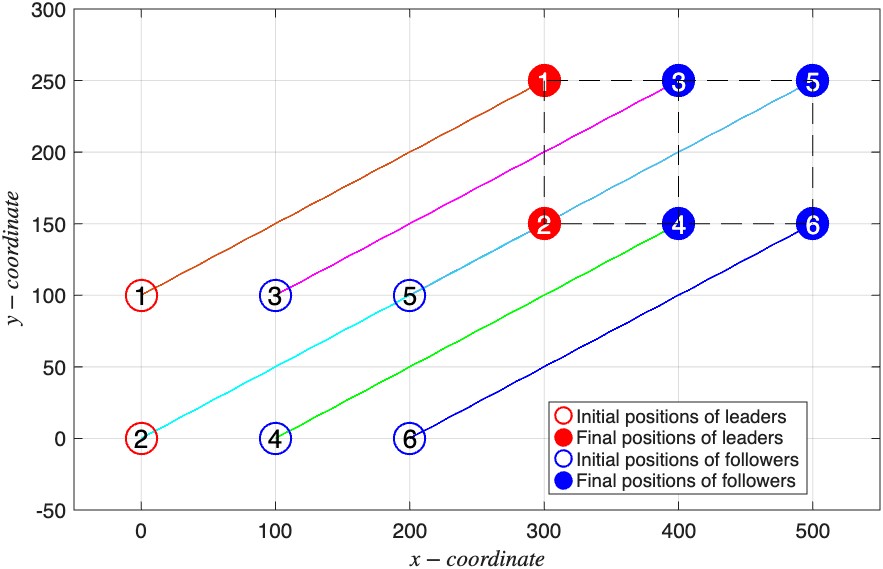}
    \caption{Formation Profile}
    \label{fig:for_profile}
    
\quad

    \includegraphics[width = 0.5\textwidth]{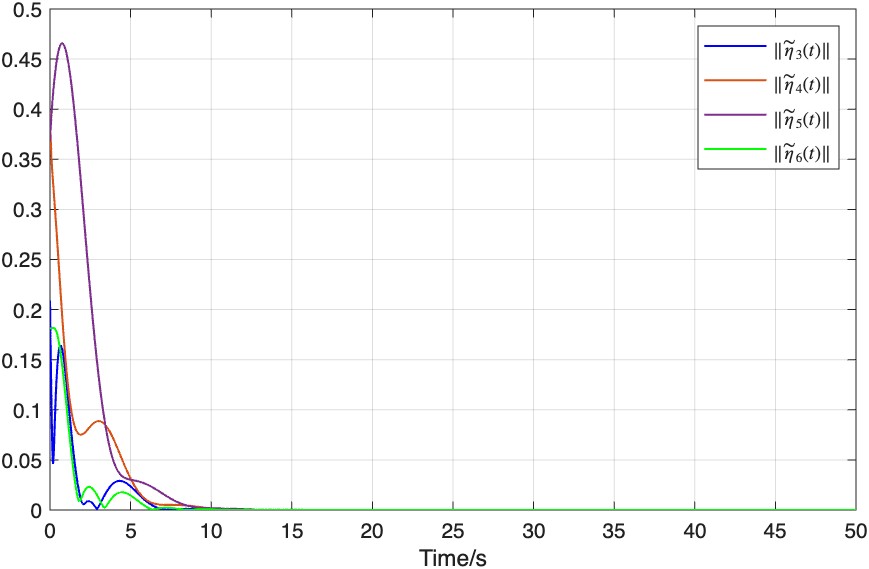}
    \caption{States of Distributed Observer}
    \label{fig:dis_obsv}

\quad

    \includegraphics[width = 0.5\textwidth]{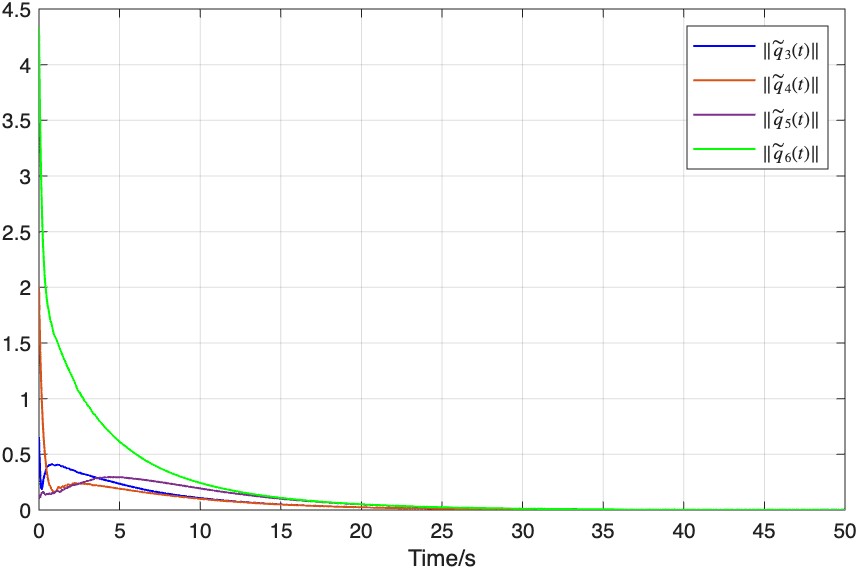}
    \caption{Distance Errors of Followers}
    \label{fig:dis_error}
\end{figure}
In this section, we implement the control law \eqref{law1} on a group of Euler–Lagrange agents operating in a planar space. The communication topology is illustrated in Fig.~\ref{fig:graph}, where red circles denote leaders and blue circles denote followers.
 
On the one hand, the leader system \eqref{ld_sys} is defined as
\begin{align}
    S &= \begin{bmatrix} 0 & 0 & 0 \\ 0 & 0 & \omega \\ 0 & -\omega & 0 \end{bmatrix},
    & F &= \begin{bmatrix} 0.6 & 0 & 0 \\ 0.3 & 0.3 & 0 \end{bmatrix}, \label{sim_eq3}
\end{align}
where \( \omega = \pi/2 \), and the initial condition is given by \( \eta(0) = \text{col}(10, 1, 0) \), indicating that the leaders follow a composite velocity comprising both constant and sinusoidal components. It can be verified from \eqref{sim_eq3} that the pair \((S, F)\) is observable. Hence, Assumptions~\ref{dec_amp} and~\ref{ld_amp} are satisfied.
%modified from \cite{he2021leader},

On the other hand, the Euler–Lagrange follower dynamics are governed by \eqref{fl_sys} with
\begin{align}
    M_{i}  &= \begin{bmatrix} 2.35 + 0.16\cos((q_{i})_{2}) & 0 \\ 0 & 0.10 \end{bmatrix}, \notag \\
    C_{i}  &= \begin{bmatrix}
        -0.16\sin((q_{i})_{2})(\dot{q}_{i})_{2} & -0.08\sin((q_{i})_{2})(\dot{q}_{i})_{1} \\
        0.08\sin((q_{i})_{2})(\dot{q}_{i})_{1} & 0
    \end{bmatrix}, \notag \\
    D_{i}  &= \textnormal{blkdiag}(0.3, 0.5), \quad\quad\quad\quad\quad\quad\quad\quad \forall i \in \mathcal{V}_f. \label{sim_eq1}
     \end{align}
Furthermore, from \eqref{sim_eq1}, we confirm that relation \eqref{id1} holds with the following parameterization:
\begin{align}
     Y_{i} & =\textnormal{blkdiag}((Y_{i})_{11}, (Y_{i})_{22}),\notag \\% \begin{bmatrix} (Y_{i})_{11} & \mathbf{0} \\ \mathbf{0} & (Y_{i})_{22} \end{bmatrix}, \notag \\
    \theta & = \text{col}(2.35, 0.16, -0.16, -0.08, 0.3, 0.1, 0.08, 0.5), \label{sim_eq2}
\end{align}
where
\begin{align*}
    (Y_{i})_{11} &= \begin{bmatrix} (\dot{\zeta}_{i})_{1} & \cos((q_{i})_{2})(\dot{\zeta}_{i})_{1} & \Omega_{Y_{i}} & (\dot{q}_{i})_{1} \end{bmatrix}, \\
    (Y_{i})_{22} &= \begin{bmatrix} (\dot{\zeta}_{i})_{2} & \sin((q_{i})_{2})(\dot{q}_{i})_{1}(\dot{\zeta}_{i})_{1} & (\dot{q}_{i})_{2} \end{bmatrix},
\end{align*}
with
%\begin{align*}
	 $\Omega_{Y_{i}} = \begin{bmatrix} \sin((q_{i})_{2})(\dot{q}_{i})_{2}(\dot{\zeta}_{i})_{1} & \sin((q_{i})_{2})(\dot{q}_{i})_{1}(\dot{\zeta}_{i})_{2} \end{bmatrix}$.\\
%\end{align*}
% 
It follows from \eqref{sim_eq1} and \eqref{sim_eq2} that both \( C_{i} \) and \(  Y_{i} \) remain uniformly bounded, provided that \( \dot{q}_{i}(t) \), \( \zeta_{i}(t) \), and \( \dot{\zeta}_{i}(t) \) are uniformly bounded over \( [0, +\infty) \). This ensures that Assumption~\ref{para_amp} holds.
 
The target formation shape is a rectangle with the desired bearings: %\eqref{de_bears}.
\begin{align*}
   & g_{31}^{*} = \text{col}(1, 0),~ g_{32}^{*} = \text{col}(\tfrac{\sqrt{2}}{2}, \tfrac{\sqrt{2}}{2}),~~~~ g_{34}^{*} = \text{col}(0, 1), \notag \\
   &g_{42}^{*}= \text{col}(1, 0), ~ g_{45}^{*} = \text{col}(\tfrac{\sqrt{2}}{2}, -\tfrac{\sqrt{2}}{2}),~~ g_{46}^{*} = \text{col}(-1, 0),\notag \\
   &  g_{35}^{*} = \text{col}(1, 0),~ g_{36}^{*} = \text{col}(-\tfrac{\sqrt{2}}{2}, \tfrac{\sqrt{2}}{2}),~~ g_{41}^{*} = \text{col}(\tfrac{\sqrt{2}}{2}, -\tfrac{\sqrt{2}}{2}),\notag \\
     %&  \notag \\     
    &  g_{52}^{*} = \text{col}(0, 1),~
    g_{56}^{*} = \text{col}(-\tfrac{2\sqrt{5}}{5}, \tfrac{\sqrt{5}}{5}),~g_{61}^{*} = \text{col}(\tfrac{2\sqrt{5}}{5}, -\tfrac{\sqrt{5}}{5}). %\label{de_bears}
\end{align*}
By computing \( \lambda_{\min}(\mathcal{B}_{ff}) = \underbrace{0.1457}_{ > 0} \), we verify that Assumption~\ref{uq_amp} is satisfied. The effectiveness of the control law \eqref{law1} is demonstrated via simulation with the following initial conditions:
\begin{align}
    \tilde{\eta}_{f}(0) &= (\text{\textnormal{rand}}(12,1) - 0.5 \times \mathbf{1}_{12}) \times 10, \notag \\
    s_{f}(0) &= (\text{\textnormal{rand}}(12,1) - 0.5 \times \mathbf{1}_{12}) \times 2, \notag \\
    \tilde{\theta}_{f}(0) &= (\text{\textnormal{rand}}(32,1) - 0.5 \times \mathbf{1}_{32}). \notag %\label{sim_int_val}
\end{align}

Let us compute:
\begin{align*}
    \inf_{t \geq 0,\, i,j \in \mathcal{V},\, i \neq j} \|q_{i}^{*}(t) - q_{j}^{*}(t)\| &= 100, \\
    \sqrt{n} \left( \|\tilde{q}_{f}(0)\| + \tfrac{\bar{D}}{\lambda_{\min}(\mathcal{B}_{ff})} \right) + \gamma &= 95.5864,
\end{align*}
where \( \gamma = 2 \). This confirms that the minimum inter-agent distance \( \gamma \) is maintained, satisfying Assumption~\ref{col_amp}.
 
The simulation results are illustrated in Figs.~\ref{fig:for_profile}--\ref{fig:dis_error}. Specifically, Fig.~\ref{fig:for_profile} shows the evolution of the formation profile. The convergence of the distributed observer states for the four followers is shown in Fig.~\ref{fig:dis_obsv}. Finally, the position errors of the four followers are depicted in Fig.~\ref{fig:dis_error}. As expected, the control law \eqref{law1} successfully achieves the desired objectives, validating its effectiveness.

\section{Conclusion}

This paper has investigated the problem of tracking formations driven by bearings for Euler-Lagrange systems. We first developed a distributed observer capable of handling multiple leaders under the bearing-based localization condition. By integrating this observer with the adaptive control technique, we have proposed a distributed control law that enables the agents to achieve the target formation without requiring prior knowledge of system parameters. Furthermore, we have derived a sufficient condition for collision avoidance based on the initial conditions of the closed-loop system.

While the current work does not account for external disturbances, future research will focus on enhancing the robustness of the proposed control design by addressing external disturbances of Euler-Lagrange systems.

%\section*{Appendixes}

%\section*{References}

\bibliographystyle{ifacconf}\label{sec6}% plainnat elsarticle-harv
\bibliography{myreferece}

 \end{sloppypar}

\end{document}